# Light controlled magnetoresistance and magnetic field controlled photoresistance in CoFe film deposited on BiFeO$_3$

B. Kundys,[1,a)] C. Meny,[1] M. R. J. Gibbs,[2] V. Da Costa,[1] M. Viret,[3] M. Acosta,[1] D. Colson,[3] and B. Doudin[1]

[1]*Institut de Physique et de Chimie des Matériaux de Strasbourg (IPCMS), UMR 7504 CNRS-ULP, 67034 Strasbourg, France*
[2]*Department of Materials Science and Engineering, University of Sheffield, Sheffield S1 3JD, United Kingdom*
[3]*Service de Physique de l'EtatCondensé, DSM/IRAMIS/SPEC, CEA Saclay URA CNRS 2464, 91191 Gif-Sur-Yvette Cedex, France*

We present a magnetoresistive—photoresistive device based on the interaction of a piezomagnetic CoFe thin film with a photostrictive BiFeO$_3$ (BFO) substrate that undergoes light-induced strain. The magnitude of the resistance and magnetoresistance in the CoFe film can be controlled by the wavelength of the incident light on the BiFeO$_3$. Moreover, a light-induced decrease in anisotropic magnetoresistance is detected due to an additional magnetoelastic contribution to magnetic anisotropy of the CoFe film. This effect may find applications in photo-sensing systems, wavelength detectors and can possibly open a research development in light-controlled magnetic switching properties for next generation magnetoresistive memory devices.

http://dx.doi.org/10.1063/1.4731201

Controlling the magnetization direction $\bar{M}$ in magnetic materials is a key issue for magnetic information recording. In spintronics, for example, there is a growing research activity aiming at finding methodologies to avoid the use of external stray magnetic fields, challenged by the miniaturization needs. Spin torque approach[1,2] and the use of electric field are attractive solutions. For instance, magnetoelectric materials can exhibit a change of magnetization triggered by an applied voltage, avoiding heating, and high power consumption.[3] Reports of piezoelectrically induced strain modifying the magnetic anisotropy through the piezomagnetic effect are also found in the literature.[4–10] However, the necessity to add electrical contacts to the memory elements and sensors make these approaches increasingly complex as the dimensions of magnetic devices continue to shrink. Alternative non-contact methods to control the magnetization vector are, therefore, of high interest, for example, using picosecond acoustic[11] and femtosecond laser pulses.[12]

Here, we propose an optical approach, where the incident light can result in manipulation of the magnetic anisotropy through the use of the photostrictive effect. In view of the recent interest in novel hybrid spintronics and straintronic devices,[13,14] this strategy can be especially rewarding from an energetic point-of-view and can potentially take advantage of the wavelength[15] and power dependence of the photostrictive effect to provide two supplementary degrees of freedom in devices control. Photostrictive materials are an extraordinary class of materials with practical remote control applications already effectively demonstrated.[16–18] The coupling with magnetic materials has never been reported, possibly due to the intrinsic slow response time of the photostriction effect (typically tens of seconds[19]). The recently found fast responsive photostrictive effect (below 0.1 s) in multiferroic BiFeO$_3$ (BFO) crystals can possibly overcome this problem.[20] A polycrystalline ferromagnetic Co$_{50}$Fe$_{50}$ thin film, known to exhibit a large piezomagnetic coefficient and significant magnetostriction[21–23] is chosen for the proof-of-principle experiment we propose here.

A 300 nm thick Co$_{50}$Fe$_{50}$ film was deposited onto the 100 $\mu$m thick BFO substrate (in a spontaneous polarization state) using conventional magnetron sputtering. The other dimensions of the film can be seen in Fig. 1. Light emitting diodes (LEDs) of several discrete center wavelengths (365, 455, 530, 660, and 940 nm) with typical 30 nm spectral widths were used as illumination source through an optical fibre. The LEDs were set to the equal power of 4.1 mW to study the wavelength dependence. The surface of the BFO substrate was 0.252 mm$^2$ and was illuminated uniformly with irradiance of 326 W/m$^2$.

As can be seen in Fig. 2(a), illuminating the BFO substrate at several different wavelengths results in a clear photoresistive effect defined as 100*(R$_{hv}$-R$_{dark}$)/R$_{dark}$ (the initial resistance of the CoFe film in darkness was 9.48 $\Omega$). Equal illumination time of 15 s has been chosen to study the wavelength dependence of the photo-elastically induced photoresistive effect. The irradiation with 365 nm light performed on a virgin (very first exposure to light) sample was found to create long time resistance change (Fig. 2(a)). We found that a virgin sample needs at least 24 h to recover its initial resistance level for a CoFe film after being irradiated at 365 nm for 15 s only. Without complete relaxation, further photoresistance was significantly diminished (Fig. 2(b)). Such relaxation was found to change when increasing the wavelength (44, 62, 21, and 9 s for 455, 530, 660, and 940 nm, respectively), being extremely long in the UV illumination range only. The wavelength dependence of the photoresistance follows closely optical absorption spectra of BFO substrate (Fig. 2(a) (inset)) measured for the similar piece of sample.[24]

---

[a)]Author to whom correspondence should be addressed. Electronic mail: kundys AT gmail (DOT) com

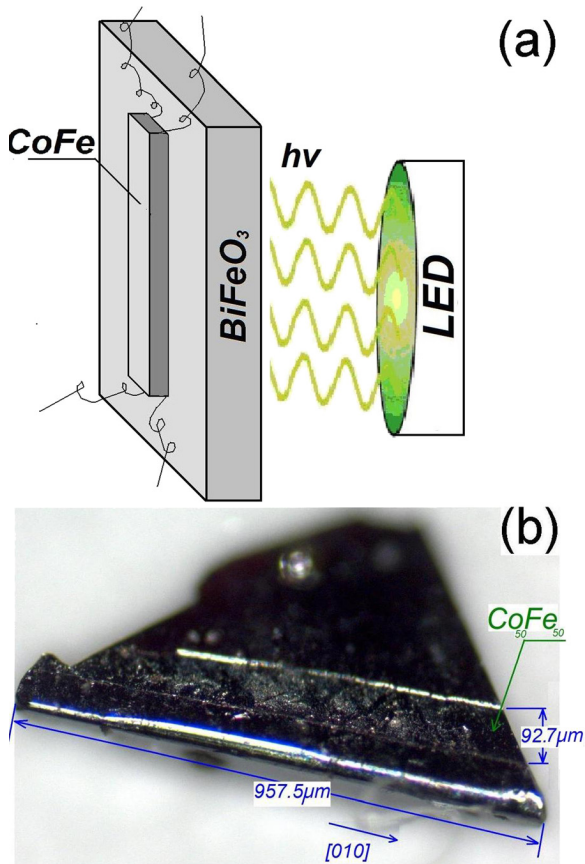

FIG. 1. (a) Schematics of the experiment. Light is illuminated on the BFO crystal along [100] direction. (b) Microscope image of the BFO with CoFe film on the top.

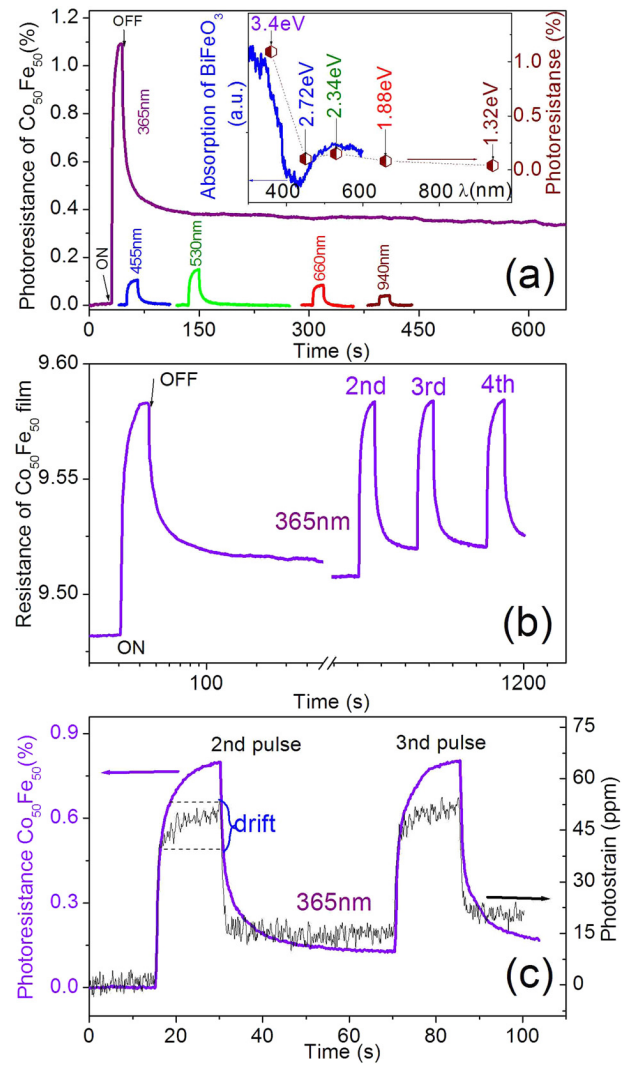

FIG. 2. (a) Time dependence of photo-induced change in resistance of CoFe film deposited on BFO substrate. Inset: (left scale) optical absorption of BiFeO$_3$ reproduced with permission from R. Moubah et al., Appl. Phys. Express 5, 035802 (2012). Copyright © 2012 The Japan Society of Applied Physics; and photoresistance (right scale). (b) Resistance of CoFe film for the first exposure to 365 nm UV light with long time relaxation and subsequent pulses. (c) Time dependence of photoresistance of CoFe film (left scale) and photostrain [101] (right scale) for 365 nm light excitation.

Generally speaking, the observed relaxation has analogy with the light induced slow capacitor charge-discharge-like mechanism that was also reported for other ferroelectric systems.[25] Similarly to the relaxation time, the magnitude of the photoresistance clearly depends on illumination wavelength (for the same light intensity) reaching ~1.1% for 365 nm light. Figure 2(c) suggests that the response/relaxation times and the magnitude of photoresistance originate from the photo-induced deformation of the BFO substrate. Indeed, the resistance of CoFe film follows closely the deformation time profile with a little delay. Photo-induced deformation was mentioned to follow photocurrent wavelength dependence in electrically polar SbSI.[26] For example, LiNbO$_3$:Fe shows two maxima in photocurrent: the first one near the bandgap energy and the second one when the light excitation energy exceeds the bandgap value.[27,28] In our sample, the same behavior can be expected. Indeed, the second maximum in photoresistance effect in CoFe (Fig. 2(a)) is observed for green light (2.34 eV) near the 2nd maximum of optical absorption and within the range of 2.3 to 3 eV bandgap values reported for BiFeO$_3$.[24,29–35]

In order to investigate possible coupling with magnetic order, we have measured the anisotropic magnetoresistance (AMR) of the CoFe film to test if the BFO strain can modify the films magnetic order. It has to be noted that AMR is known to be highly strain sensitive in other magnetic materials.[36] Experiments were performed under a saturating magnetic field of 0.5 T, rotating the field in the plane of the sample of Fig. 1, comparing data in the dark and under LED illumination (Fig. 3(a)). The resistance is maximal when the direction of current is parallel to the applied magnetic field and directed along the longest dimension of the film. Due to the substrate mediated light-induced deformation, a magnetoelastic anisotropy appears in the CoFe film and decreases the AMR. The magnitude depends on wavelength, following the trend of Fig. 2(a).

This observation is further supported by R(H) measurements at fixed angle (Fig. 3(b)). The magnetoresistance becomes smaller under light when the magnetic field is perpendicular to the longest dimension of the CoFe film (and current Fig. 3(b)), in contrast to the observed superposed curves in the parallel configuration (Fig. 3(c)). Note that the shape of the curves confirms that 0.5 T was enough to saturate the film magnetization.

To explain this effect, one has to consider the dependence of the magnetoresistance on the angle between the direction of electrical current and orientation of the magnetization in the magnetic material. In the macroscopic approach, the light-induced deformation applied to the CoFe film changes the distribution of magnetic moments via the inverse magnetostrictive effect (Villari effect). Due to the magnetoelastic contribution to the magnetic anisotropy, it becomes energetically costly to change the magnetization orientation by a magnetic field. Because magnetoresistance depends on the ability of magnetic moments to change their orientation in response to magnetic field, the overall magnetoresistance decreases. More interestingly, we also found that the inverse effect can be observed, i.e., modifying the photoresistance magnitude by an external magnetic field (Fig. 3(c)).

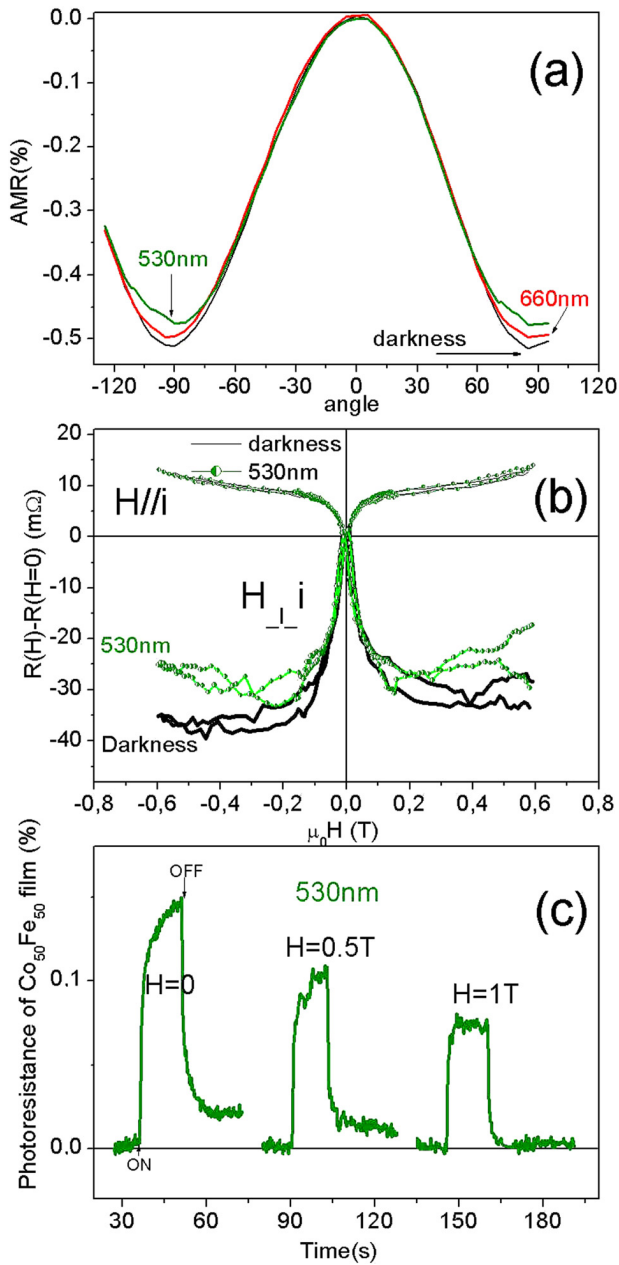

FIG. 3. (a) AMR of the CoFe film at 0.5 T under green (530 nm) and blue (455 nm) illumination and in darkness, respectively. (b) Magnetoresistance loops in darkness and under 530 nm light with magnetic field in perpendicular and in parallel configuration. (c) Photoresistance at different magnetic field values in the perpendicular configuration.

To further explore the origins of the photoresistive behaviour, we have also measured the temperature of the sample directly during illumination (Fig. S1(a)). Details of the results are given in the supplementary material.[37] The Pt thermometer attached to the sample during illumination has detected a temperature change not exceeding $(1.5 \pm 0.1)$ K under 365 nm (largest energy) illumination for 15 s (Fig. 2). The related thermo-induced resistance increase (Fig. S1(b)) does not exceed $(0.21 \pm 2)$%. This shows that the long time scale relaxation (Fig. 2) is not related to thermal heating of CoFe film by light. More generally, the photoresistance response is much faster than the temperature change of the sample.[37] A similar temperature analysis of the $BiFeO_3$ substrate heating under light also allowed us to discard a model of dominant thermo-elongation of the substrate.[38] This effect is nevertheless a possible source of the measured elongation drift after saturation (Fig. 2(c)) and the change in the slope in corresponding photoresistance effect.

In summary, we reported a proof-of-concept experiment showing the possible photoresistive coupling and optical control of the magnetic anisotropy in the CoFe/BFO. We find that deformation of BFO substrate under illumination with visible light can modify the resistance and AMR values of a ferromagnetic film though the photoelastic effect. The magnitude of both the photoresistance and AMR clearly depends on wavelength. Beside light-controlled resistance or applications in other photo-sensing systems, it can be anticipated that this finding can possibly have potential applications in light-controlled magnetic switching properties for the next-generation of magnetoresistive memory devices. Although the magnitude of both the reported photo-resistive and opto-magnetic couplings is small, further improvements can be expected using substrates with larger photostriction and optimal thicknesses of the both components: substrate[39] and piezomagnetic film. Therefore, a novel trend of research in this area might be opened in which multi-functional spintronic devices properties can be envisioned in inorganic hybrid straintronic-spintronic structures.

# Supplementary Material

The Pt thermometer attached to the sample has detected (1.5±0.1)K temperature change of the sample (fig. S1a) under 365nm light of largest energy for 15s illuminating time period (sufficient time to saturate the photoresistance of CoFe film (fig. 2)).

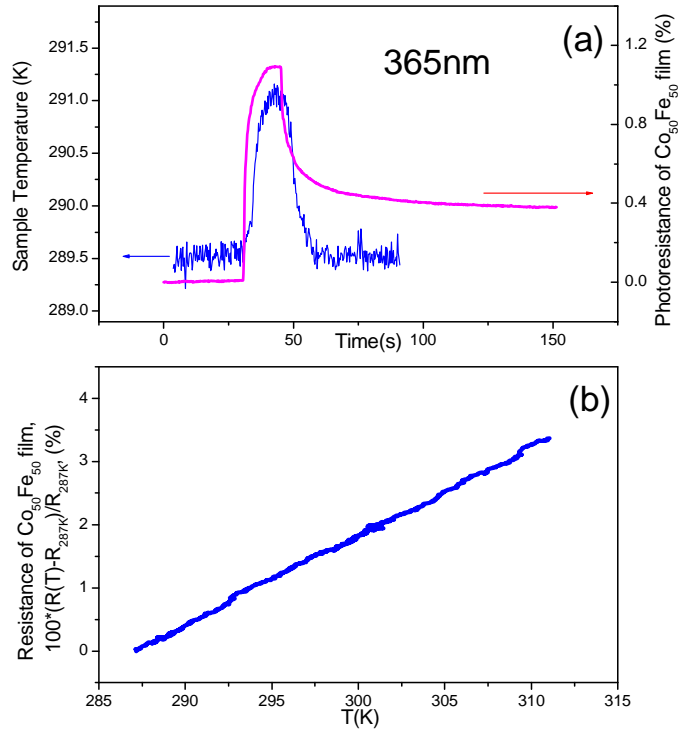

FIG. S1. (Color online) (a) Time dependence of the photoresistance (right scale) and sample temperature change(left scale) due to light 365nm. (b) Temperature dependence of the resistance of CoFe film measured with external heating source.

The time scale also shows that photoresistance response is much faster than the temperature change of the sample. Furthermore, measured resistance of CoFe film as a function of temperature (fig. S1b) showed that the heating of CoFe film by 1.5K results in the resistance increase by (0.21±2)%. Same procedure has been performed for other wavelengths. Corresponding values (found not to exceed 23% of the total effect) should be subtracted from reported photoresistance to get real photoelastic-piezomagnetic coupling without thermal contribution coming from indirect CoFe heating.